\documentclass[journal]{IEEEtran}
\pdfoutput=1

\usepackage{graphicx}

\usepackage{siunitx}
\usepackage[caption=false]{subfig}

\usepackage{textcomp}

\usepackage{hyperref}

\begin{document}

% <><><><><><><><><><><><><><><><><><><><><><><><><><><><><><><><><><><><><><><>
% Title page
% <><><><><><><><><><><><><><><><><><><><><><><><><><><><><><><><><><><><><><><>
\title{The ONSEN Data Reduction System for the Belle~II Pixel Detector}

\author{Thomas~Ge\ss{}ler,
        Wolfgang~K\"u{}hn,~\IEEEmembership{Member,~IEEE,}
        Jens~S\"o{}ren~Lange,
        Zhen'An~Liu,~\IEEEmembership{Senior Member,~IEEE,}
        David~M\"u{}nchow,
        Bj\"o{}rn~Spruck,
        and~Jingzhou~Zhao,~\IEEEmembership{Student Member,~IEEE}%
\thanks{\textcopyright 2014, 2015 IEEE. Personal use of this material is permitted.
        Permission from IEEE must be obtained for all other uses, in any
        current or future media, including reprinting/republishing this
        material for advertising or promotional purposes, creating new
        collective works, for resale or redistribution to servers or lists, or
        reuse of any copyrighted component of this work in other works.}%
\thanks{This is the author's version of an article that has been published in
        IEEE Transactions on Nuclear Science. Changes were made to this version
        by the publisher prior to publication. The final version of record is
        available at \url{http://dx.doi.org/10.1109/TNS.2015.2414713}.
        Manuscript received June 16, 2014; revised November 25, 2014; accepted
        March 10, 2015. Manuscript received in final form on March 17, 2015.
        Date of publication June 12, 2015.}%
\thanks{This work was supported by the German Bundesministerium f{\"u}r Bildung
        und Forschung under grant number 05H12RG8. The research leading to
        these results has received funding from the European Commission under
        the FP7 Research Infrastructures project AIDA, grant agreement number
        262025.}%
\thanks{T.~Ge\ss{}ler, W.~K\"u{}hn, J.~S.~Lange, D.~M\"u{}nchow, and B.~Spruck
        are with the II.\ Physikalisches Institut, Justus Liebig University
        Giessen, 35392 Gie\ss{}en, Germany (e-mail:
% With default hyphenation, the e-mail adresses are printed in a horrible-
% looking way. We make some overrides:
%  - allow breaks after the @;
%  - allow breaks after the TLD elements (after .), but:
%     * not before the @, because this would produce a seemingly complete, but
%       incorrect e-mail address on the next line; and
%     * not before .de, because this would look stupid; also:
%  - forbid breaking uni-giessen, since this should be one element.
        Thomas.Gessler@\allowbreak{}exp2.\allowbreak{}physik.\allowbreak{}%
        \mbox{uni-giessen}.de;
        Wolfgang.Kuehn@\allowbreak{}exp2.\allowbreak{}physik.\allowbreak{}%
        \mbox{uni-giessen}.de;
        Soeren.Lange@\allowbreak{}exp2.\allowbreak{}physik.\allowbreak{}%
        \mbox{uni-giessen}.de;
        David.Muenchow@\allowbreak{}exp2.\allowbreak{}physik.\allowbreak{}%
        \mbox{uni-giessen}.de;
        Bjoern.Spruck@\allowbreak{}exp2.\allowbreak{}physik.\allowbreak{}%
        \mbox{uni-giessen}.de%
        ).}%
\thanks{Z.~Liu and J.~Zhao are with the Institute of High Energy Physics,
        Chinese Academy of Sciences, 100049 Beijing, China (e-mail:
        liuza@ihep.ac.cn;
        zhaojz@ihep.ac.cn%
        ).}%
}

\maketitle
\thispagestyle{empty}
\pagestyle{empty}

% <><><><><><><><><><><><><><><><><><><><><><><><><><><><><><><><><><><><><><><>
% Abstract and keywords
% <><><><><><><><><><><><><><><><><><><><><><><><><><><><><><><><><><><><><><><>
\begin{abstract}
    We present an FPGA-based online data reduction system for the pixel
    detector of the future Belle~II experiment. The occupancy of the pixel
    detector is estimated at \SI[detect-weight]{3}{\percent}. This corresponds
    to a data output rate of more than
    \SI[per-mode=symbol,detect-weight]{20}{\giga\byte\per\second} after zero
    suppression, dominated by background. The \emph{Online Selection Nodes}
    (ONSEN) system aims to reduce the background data by a factor of 30. It
    consists of 33 MicroTCA cards, each equipped with a Xilinx Virtex-5 FPGA
    and \SI[detect-weight]{4}{\gibi\byte} DDR2 RAM. These cards are hosted by 9
    AdvancedTCA carrier boards. The ONSEN system buffers the entire output data
    from the pixel detector for up to 5 seconds. During this time, the Belle~II
    high-level trigger PC farm performs an online event reconstruction, using
    data from the other Belle~II subdetectors. It extrapolates reconstructed
    tracks to the layers of the pixel detector and defines regions of interest
    around the intercepts. Based on this information, the ONSEN system discards
    all pixels not inside a region of interest before sending the remaining
    hits to the event builder system. During a beam test with one layer of the
    pixel detector and four layers of the surrounding silicon strip detector,
    including a scaled-down version of the high-level trigger and data
    acquisition system, the pixel data reduction using regions of interest was
    exercised. We investigated the data produced in more than 20 million events
    and verified that the ONSEN system behaved correctly, forwarding all pixels
    inside regions of interest and discarding the rest.
\end{abstract}

\begin{IEEEkeywords}
    Data acquisition, field programmable gate arrays, high energy physics
    instrumentation computing.
\end{IEEEkeywords}

% <><><><><><><><><><><><><><><><><><><><><><><><><><><><><><><><><><><><><><><>
% Main matter
% <><><><><><><><><><><><><><><><><><><><><><><><><><><><><><><><><><><><><><><>

% ------------------------------------------------------------------------------
\section{Introduction}
\begin{figure}
    \centering
    \includegraphics[width=\linewidth]{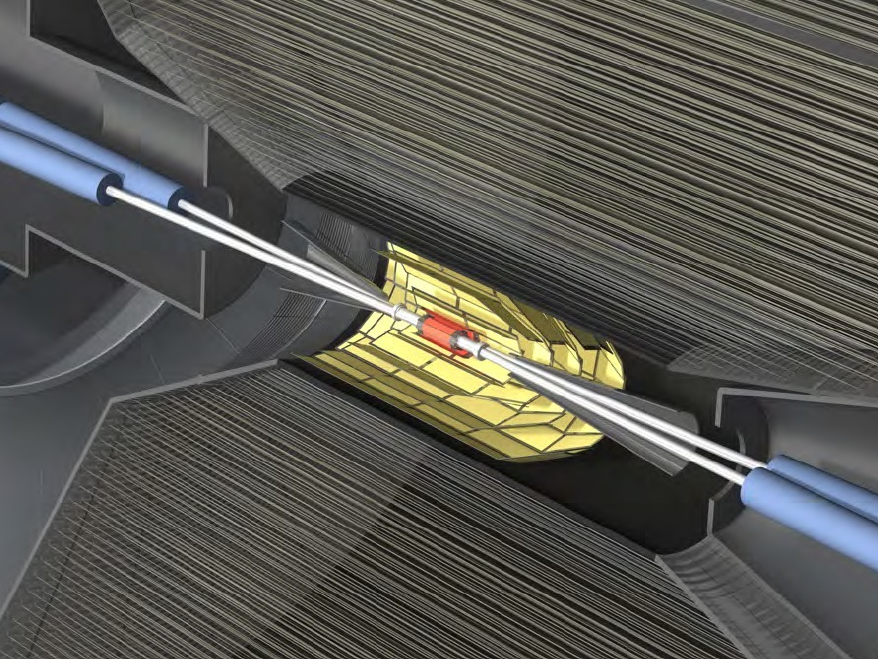}
    \caption{\label{fig:belle2ir}%
        Cutaway drawing of the Belle~II interaction region. The visible
        subdetectors, from the interaction points radially outwards, are: the
        pixel detector (PXD, two barrel layers), the silicon vertex detector
        (SVD, four barrel layers), and the central drift chamber (CDC, with
        wires shown).%
    }
\end{figure}

\IEEEPARstart{O}{ne} of the goals of the Belle~II
experiment~\cite{2010arXiv1011.0352A} is the improvement of the CP-violation
parameter measurements made by its predecessor Belle. This makes a very precise
determination of particle decay vertices necessary. For vertex measurement,
Belle used four layers of double-sided silicon strip detectors in a barrel
arrangement around the interaction point, called \emph{silicon vertex detector}
(SVD). In order to achieve a higher vertex resolution than Belle, Belle~II will
feature two additional layers of DEPFET~\cite{Richter2003250} \emph{pixel
detectors} (PXD) very close to the beam pipe, surrounded by four SVD layers.
This arrangement is shown in figure \ref{fig:belle2ir}.

The inner PXD layer will have a radius of only \SI{14}{mm}. As a consequence of
this very close proximity to the interaction point, the PXD will detect many
hits from background events, mainly two-photon QED processes. The number of
fired pixels per read-out cycle (\SI{20}{\micro\second}) from such processes
will be in the order of \SI{100000}{} for the whole detector, while only a
small number of hits will belong to charged tracks from events relevant to the
Belle~II physics program. It is necessary to extract these signal hits from the
much larger background, because the Belle~II event builder system cannot cope
with the PXD output data rate of over
\SI[per-mode=symbol]{20}{\giga\byte\per\second}---more than 10 times that of
all other subdetectors combined. With the PXD data alone, however, it is
virtually impossible to distinguish signal hits from background hits.

\begin{figure*}
    \centering
    \includegraphics[width=\linewidth]{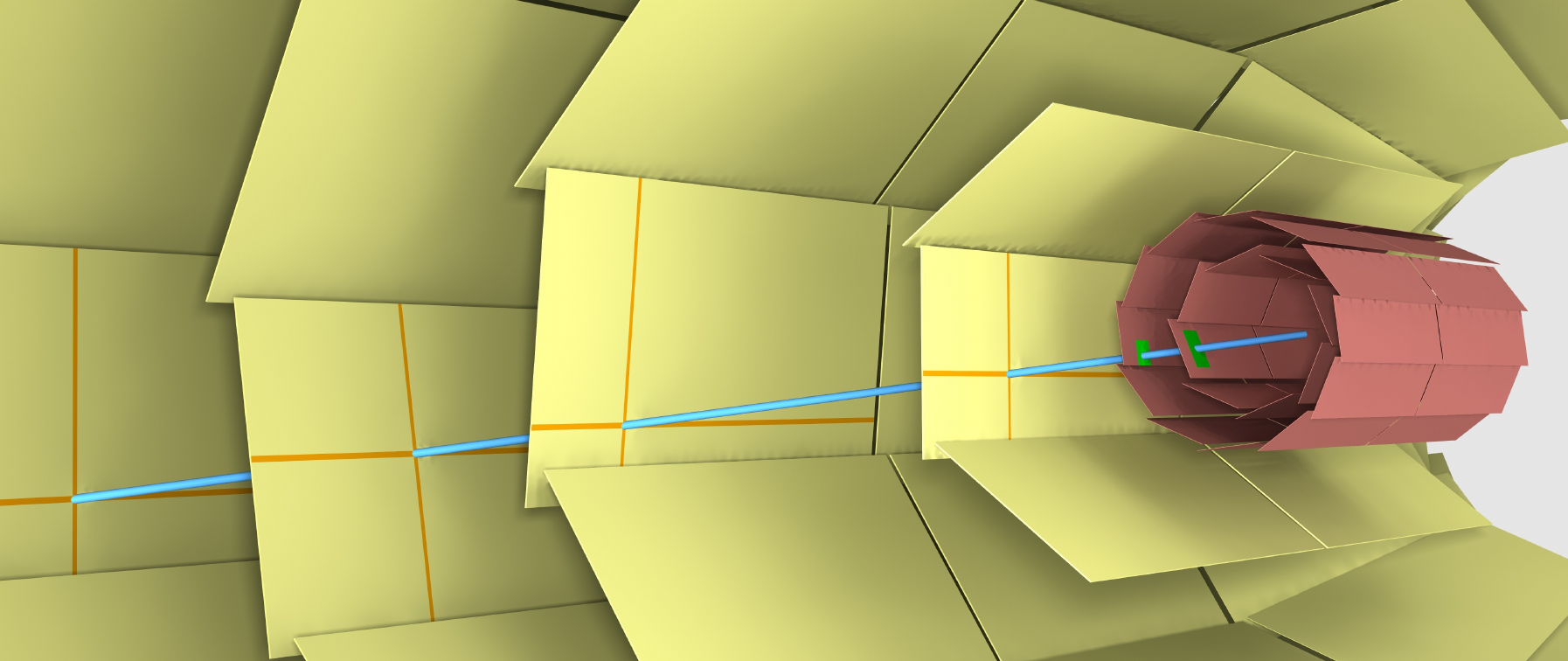}
    \caption{\label{fig:render_roi}%
        Illustration of the ROI generation mechanism. The active areas of the
        PXD and half of the SVD are shown. The HLT and the DATCON use SVD strip
        hits to reconstruct a particle track. The track is extrapolated to the
        PXD, and regions of interest are defined around the intercepts with the
        detector layers.%
    }
\end{figure*}
A distinction between background and signal becomes possible when data from the
outer tracking detectors are taken into account: Particles produced in
background processes typically have a low transversal momentum; their tracks
reach the PXD, but not the SVD or the central drift chamber (CDC) surrounding
it. The tracks from most relevant events, however, have a momentum high enough
to reach at least the SVD. By reconstructing these tracks and finding their
intercept with the layers of the PXD, rectangular areas can be determined,
inside of which the corresponding hits on the PXD layers are expected. These
areas are referred to as \emph{regions of interest} (ROIs). The intercept of an
extrapolated track with one of the PXD layers defines the center of an ROI on
that layer. The ROI size depends on the uncertainty of the track reconstruction
and extrapolation processes. The concept of ROI generation is illustrated in
figure~\ref{fig:render_roi}.

% ------------------------------------------------------------------------------
\section{Data Acquisition and ROI Generation}
\begin{figure}
    \centering
    \includegraphics[width=\linewidth]{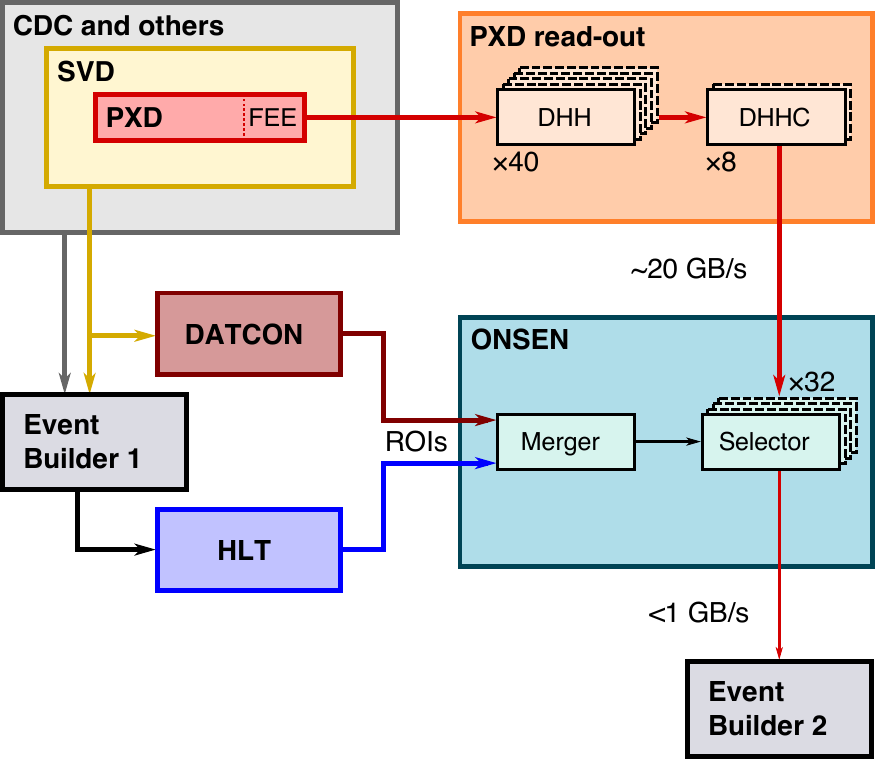}
    \caption{\label{fig:setup_belle2}%
        Simplified PXD read-out and ROI generation scheme for the Belle~II
        experiment. Only the subsystems relevant for PXD data reduction are
        shown.%
    }
\end{figure}%
The data path for the PXD data and ROIs in Belle~II is shown in
figure~\ref{fig:setup_belle2}. Each of the 40 PXD modules---so-called
\emph{half-ladders}---includes several ASICs as front-end electronics. They
digitize the charge of each pixel, convert it to a zero-suppressed format, and
store the digitized data in a buffer. Upon receiving a hardware trigger, a
\emph{Data Handling Hybrid} (DHH)~\cite{dhh} for each half-ladder reads these
data out. The DHH is an FPGA board that is also responsible for initializing
the ASICS of the half-ladder and setting configuration parameters, such as
pedestal and threshold values. For every 5 DHHs, a \emph{DHH Controller} (DHHC)
performs a 5-to-4 multiplexing: It merges the data from 2 half-ladders of the
inner PXD layer and 3 half-ladders of the outer PXD layer into a subevent (see
figure~\ref{fig:dist_dhh}). Each of the 8 DHHCs then sends the merged data to
the ONSEN system (see below) on one of its four output links, alternating
between links on an event-by-event basis. This mechanism compensates for the
higher occupancy closer to the beam pipe and effectively balances the load
across the output links. The total number of output links from the PXD read-out
system is 32.

\begin{figure}
     \centering%
     \subfloat[][]{%
        \includegraphics[width=\linewidth]{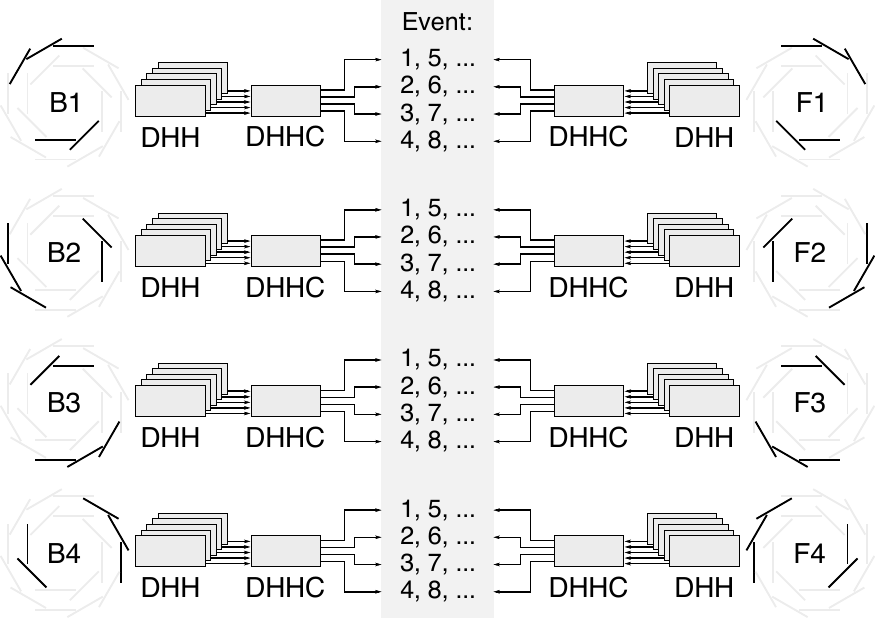}%
        \label{fig:dist_dhh}}\\%
     \subfloat[][]{%
        \includegraphics[width=\linewidth]{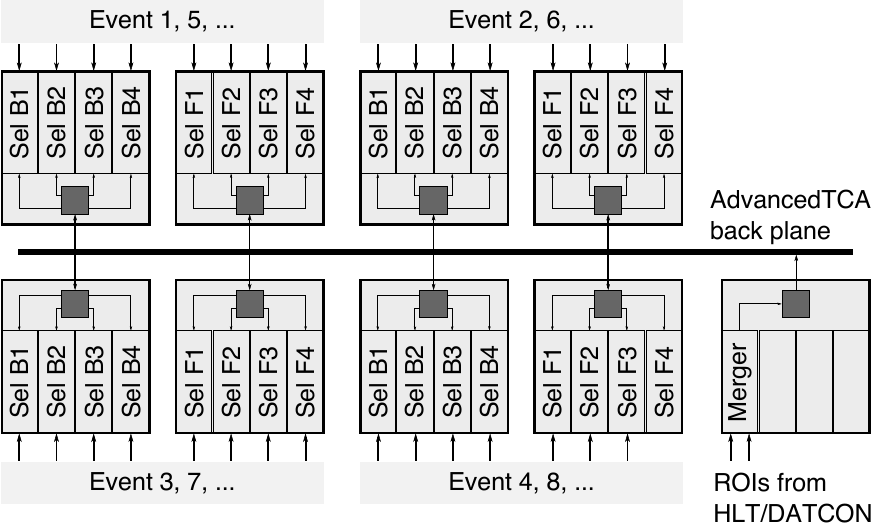}%
        \label{fig:dist_onsen}}%
     \caption{\label{fig:dist}%
         Conceptual illustration of the PXD data path.
         \protect\subref{fig:dist_dhh}
         The 40 PXD half-ladders are divided into eight sections: four in
         backward (B) and four in forward (F) direction. One DHHC handles the
         data from each section and distributes them between four outputs,
         based on the event number.
         \protect\subref{fig:dist_onsen}
         Each of the 32 ONSEN selector cards receives the data from one of the
         eight PXD sections for one out of every four events.%
     }
\end{figure}
Two independent subsystems of the Belle~II DAQ chain create ROIs for an event:
\begin{enumerate}
    \item
    For events selected by the hardware triggers, the \emph{Event Builder 1}
    creates subevents with data from all detectors \emph{except} the PXD. With
    this information, the \emph{high-level trigger} (HLT)---a highly parallel
    PC farm---performs an online event reconstruction and makes a physics-level
    event selection (software trigger)~\cite{6578610}. The HLT extrapolates the
    tracks that were found during the event reconstruction step to the PXD and
    uses them for ROI generation. Depending on the event topology, a latency of
    up to 5 seconds is possible between the hardware trigger for an event and
    the response from the HLT, including software trigger and ROIs.
    Furthermore, the HLT output is unordered: Since the HLT processes many
    events in parallel, a recent event with a simple structure may already be
    finished, while an older, more complicated event is still being processed.
    \item
    The \emph{data concentrator} (DATCON) is an FPGA-based online track
    reconstruction system that operates on the SVD data. The hardware platform
    is, for the most part, identical to that of the ONSEN system (see below).
    The tracking is based on a two-step approach: a sector-neighbor finder and
    a fast Hough transformation. The DATCON extrapolates reconstructed tracks
    and uses them to determine ROIs on the PXD planes, similar to the HLT.
    However, since it processes events in a pipelined fashion on FPGAs, the
    maximum processing time per event is around \SI{10}{\micro\second}, and the
    output is ordered.
\end{enumerate}
The existence of two separate ROI sources and the high latency of the HLT
output make a system necessary that:
\begin{enumerate}
    \item
    receives and buffers the data output from the PXD;
    \item
    receives and buffers the ROIs generated by the \mbox{DATCON};
    \item
    receives software triggers and ROIs from the HLT;
    \item
    matches all ROIs and data for an event as soon as the HLT ROIs have arrived;
    and
    \item
    performs the actual pixel data reduction, based on the combined ROIs from
    both sources.
\end{enumerate}
The hardware platform for such a system must meet certain requirements,
including: fast optical serial links for the reception of PXD data; gigabit
Ethernet interfaces for both the reception of the software trigger and ROIs and
the output of the reduced data; large memory, allowing the buffering of the PXD
data; high memory and I/O bandwidth; and high-speed data processing
capabilities.

% ------------------------------------------------------------------------------
\section{The ONSEN System} The \emph{Online Selection Nodes} (ONSEN) system was
developed to fulfill the requirements stated above. An earlier version of the
system was described in a previous publication~\cite{6615996}. The hardware
platform has since been upgraded, and many changes were made to data formats,
implementation specifics, and the overall data flow concept. The new system was
tested extensively in both laboratory and beam tests.

\subsection{Hardware}
\begin{figure}
     \centering%
     \subfloat[][]{%
        \includegraphics[width=\linewidth]{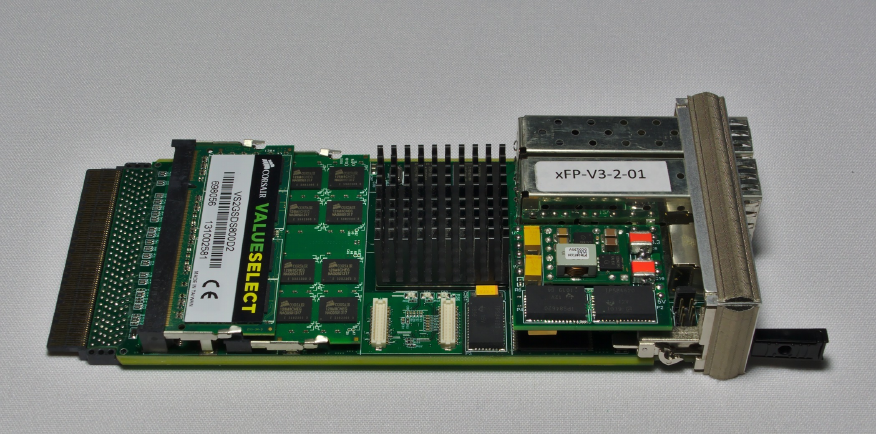}%
        \label{fig:hw_xfp}}\\%
     \subfloat[][]{%
        \includegraphics[width=\linewidth]{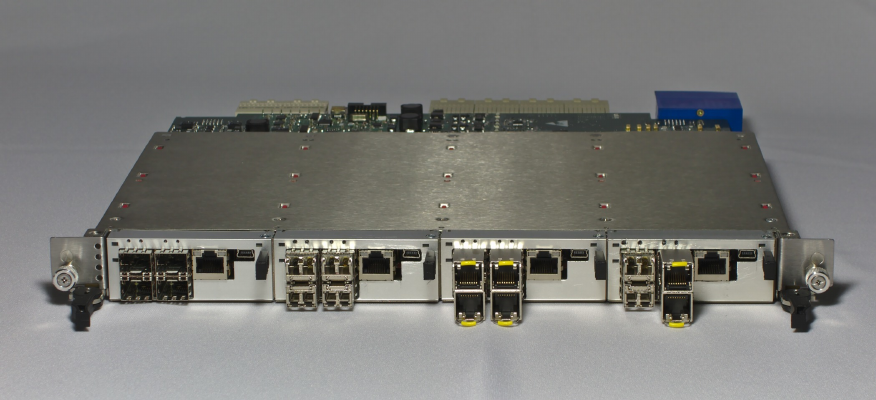}%
        \label{fig:hw_carrier}}%
     \caption{\label{fig:hw}%
         Hardware used for the ONSEN system.
         \protect\subref{fig:hw_xfp}
         An xFP card with a \mbox{Virtex-5} FPGA. The final system will use 33
         cards of this type.
         \protect\subref{fig:hw_carrier}
         A ``Compute Node'' carrier board, populated with four xFP cards. The
         cards are equipped with SFP+ transceivers for either optical fibers or
         RJ45 Ethernet cables. The final system will include 9 such carrier
         boards.%
     }
\end{figure}
The hardware platform for the ONSEN system will consist of 33 \emph{xTCA-based
FPGA Processor} (xFP) cards~\cite{xfp} hosted by 9 AdvancedTCA carrier boards.
It is a development of IHEP Beijing in collaboration with JLU Giessen.
Figure~\ref{fig:hw_xfp} shows an xFP card of the third hardware revision. This
is the final version, which will be used in the Belle~II experiment. It
features a Xilinx Virtex-5 FX70T FPGA with an embedded PowerPC CPU. Two DDR2
modules provide \SI{4}{\gibi\byte} of RAM---enough to store the output of one
DHHC for at least five seconds. An RJ45 port, used with the embedded Ethernet
MAC of the Virtex-5, provides gigabit Ethernet connectivity. Four SFP+ cages
connect to multi-gigabit serial transceivers of the FPGA. They can be equipped
with different pluggable modules, each adding either an optical fiber
interface, running at up to \SI[per-mode=symbol]{6.25}{\giga\bit\per\second},
or an additional RJ45 socket for gigabit Ethernet. Six more multi-gigabit
transceivers are routed to the module's AMC connector.

Figure~\ref{fig:hw_carrier} shows the ``Compute Node'' carrier board. It hosts
four xFP cards and establishes a full-mesh interconnection between them, using
the multi-gigabit links on the AMC connectors. In addition, the carrier board
features a \mbox{Virtex-4} FX60 FPGA that connects to the AdvancedTCA back
plane with 16 multi-gigabit links and to each xFP card with four duplex LVDS
ports. This FPGA acts as a switch: When multiple carrier boards, each equipped
with one or more xFP cards, are used together in an AdvancedTCA shelf, the
switch FPGAs allow the xFP cards to communicate with each other. Figure
\ref{fig:dist_onsen} shows how these channels will be employed in the ONSEN
system for the distribution of ROIs in the system. This mechanism will be
explained in more detail below.

\subsection{Implementation}
Using the information from the HLT, the ONSEN system achieves the required data
reduction factor of 30 in two steps: A positive software trigger decision is
expected for about one-third of all events. If the HLT rejects an event, the
ONSEN system discards all corresponding pixel data. This happens regardless of
possible ROIs found by the DATCON, since the data from the other subdetectors
is lost in case of a rejected trigger. By this mechanism, the software trigger
reduces the average trigger rate from \SI{30}{\kilo\hertz} to
\SI{10}{\kilo\hertz}, effectively dividing the output data rate by 3.

The ROIs for accepted events will cover approximately 1/10 of the sensitive PXD
area. Since most PXD hits stem from background events, the ONSEN system reduces
the output data rate by another factor of 10 by discarding hits that are not
inside ROIs.

Two pairs of FPGA configurations for the 33 xFP cards and their associated
carrier boards will be used in the ONSEN system:
\begin{enumerate}
    \item
    One of the 33 xFP cards acts as the \emph{merger node}. It receives ROIs
    from the DATCON via optical fiber and stores them in RAM. A pointer to the
    stored information, along with the corresponding event number, is written
    to a look-up table. The merger node then waits for the HLT to send a
    software trigger decision, and possibly ROIs, for the same event. The
    connection between HLT and merger node is established via gigabit Ethernet,
    using an RJ45 SFP+ transceiver. As transport layer protocol, an FPGA
    implementation of either TCP (namely SiTCP~\cite{4545224}) or UDP will be
    used.

    Upon receiving the software trigger decision and HLT ROIs for an event, the
    merger node reads back the corresponding DATCON ROIs from memory. It
    combines the ROIs from both sources into a single packet and transfers it
    to the switch FPGA on the carrier board.

    The switch FPGA uses the back plane of the \mbox{AdvancedTCA} shelf to
    distribute the ROI packets to the other boards in the ONSEN system, which
    perform the actual selection of the pixel data. The switch FPGA performs a
    preselection: It only distributes the relevant ROI packets (i.e., those for
    the associated event numbers) to each of the other carrier boards. The
    preliminary mapping of PXD sections and event numbers to the individual
    boards in the ONSEN system is shown in figure~\ref{fig:dist_onsen}.

    \item
    The remaining 32 xFP cards are the \emph{selector nodes}. They populate 8
    carrier boards in groups of 4. Each selector node is connected to one of
    the output links of a DHHC via a
    \SI[per-mode=symbol]{6.25}{\giga\bit\per\second} optical fiber link. It
    receives the data from one of the eight PXD sections for one out of every
    four events. Eight selector nodes cover all sections of the PXD, so that
    the cards hosted by two carrier boards process the data for a complete
    event.

    When a selector node receives the PXD data for an event, it writes them to
    RAM and stores the pointer and event number in a look-up table (much like
    the merger node does for the DATCON ROIs). After no more than 5 seconds,
    the merger node will have received and processed an HLT packet for the same
    event, and its switch FPGA will have distributed it over the
    \mbox{AdvancedTCA} back plane. The other switch FPGAs receive these packets
    and distribute them to the selector nodes. When the ROIs for an event
    arrive, the selector nodes read back the relevant pixel data from memory,
    filter them according to the ROI information, and send the reduced data to
    the Belle~II Event Builder 2. For this connection, again, an RJ45 SFP+
    transceiver and either SiTCP or UDP are used.
\end{enumerate}
The logic for both merger and selector nodes has been implemented in VHDL and
tested extensively. A Linux system, running on the embedded CPUs in the FPGAs
of the xFP cards and carrier boards, is used for debugging, monitoring, and
online configuration of the ONSEN system. It can be accessed over gigabit
Ethernet via the RJ45 ports on the front panel of the xFP cards. Slow control
functions of the FPGA logic are made available through registers on an embedded
bus system. They are accessible via an EPICS daemon~\cite{epics} running on the
Linux system. This configuration allows the integration of the ONSEN system
into the overall Belle~II slow control scheme.

% ------------------------------------------------------------------------------
\section{Beam Test in January 2014}
\begin{figure}
    \centering
    \includegraphics[width=\linewidth]{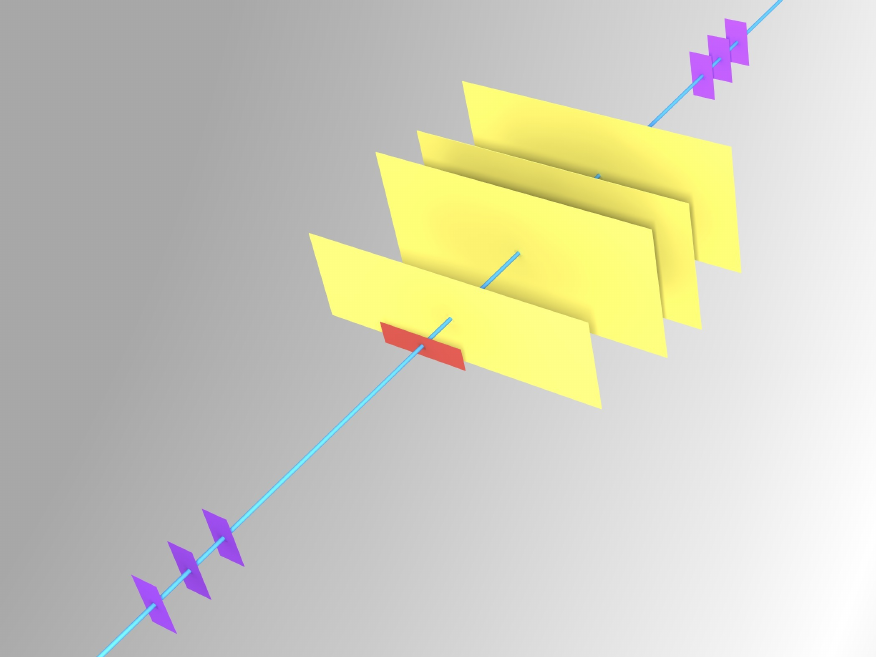}
    \caption{\label{fig:render_tb}%
        Detector setup for the beam test at DESY in January 2014; from front to
        back: 3 EUDET layers, 1 PXD layer, 4 SVD layers, and 3 more EUDET
        layers%
    }
\end{figure}%
During a four-week test at the DESY test beam facility in January 2014, modules
of the Belle~II PXD and SVD were used in conjunction for the first time. The
setup for this test included a ``radial slice'' of the vertex detector: one PXD
layer and four SVD layers, allowing to exercise data acquisition with both
systems as well as online track reconstruction, ROI generation, data reduction,
and event building. Figure~\ref{fig:render_tb} shows the detector configuration
that was used during most runs: Beside the PXD and SVD modules, six layers of
the EUDET pixel detector~\cite{Rubinskiy2012923}, available at the DESY test
beam facility, were included as a ``beam telescope'' for alignment and
verification purposes.

The momentum and rate of the electron or positron beam available in the test
beam area could be varied by choosing different conversion targets, currents
for the beam spread magnet, and collimator settings. Momenta of up to
\SI[per-mode=symbol]{6}{\giga\electronvolt\per c} and rates of up to
\SI[per-mode=symbol]{6}{\kilo\hertz} were available.

\begin{figure}
    \centering
    \includegraphics[width=\linewidth]{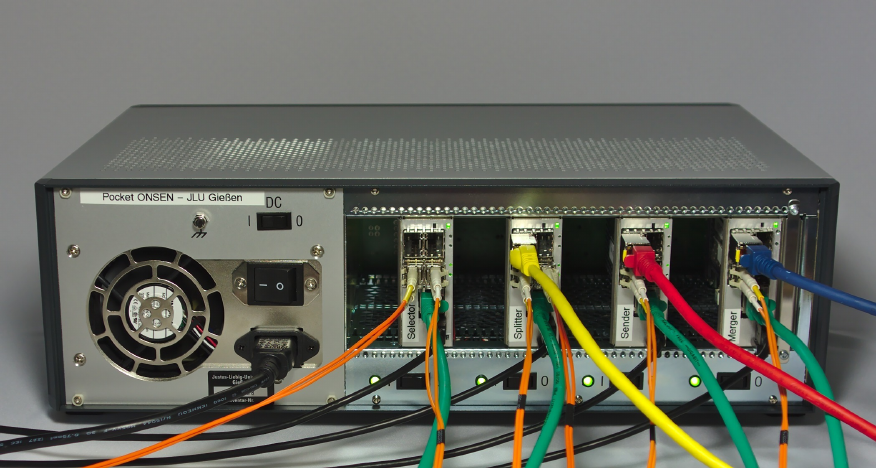}
    \caption{\label{fig:hw_pocket_onsen}%
        Pocket ONSEN system, as used during the beam test: four xFP cards in a
        MicroTCA shelf, with back plane interconnection used for ROI transfer%
    }
\end{figure}
A scaled-down but fully functional ``Pocket DAQ'' system was set up for this
test, including the HLT and event builder systems. Similarly, a ``Pocket
ONSEN'' system was used to exercise the data reduction mechanism (see figure
\ref{fig:hw_pocket_onsen}): Instead of an AdvancedTCA shelf with carrier
boards, a MicroTCA shelf with individual xFP cards was used, allowing direct
connections between the boards through the back plane without an intermediate
switch.

\begin{figure}
    \centering
    \includegraphics[width=\linewidth]{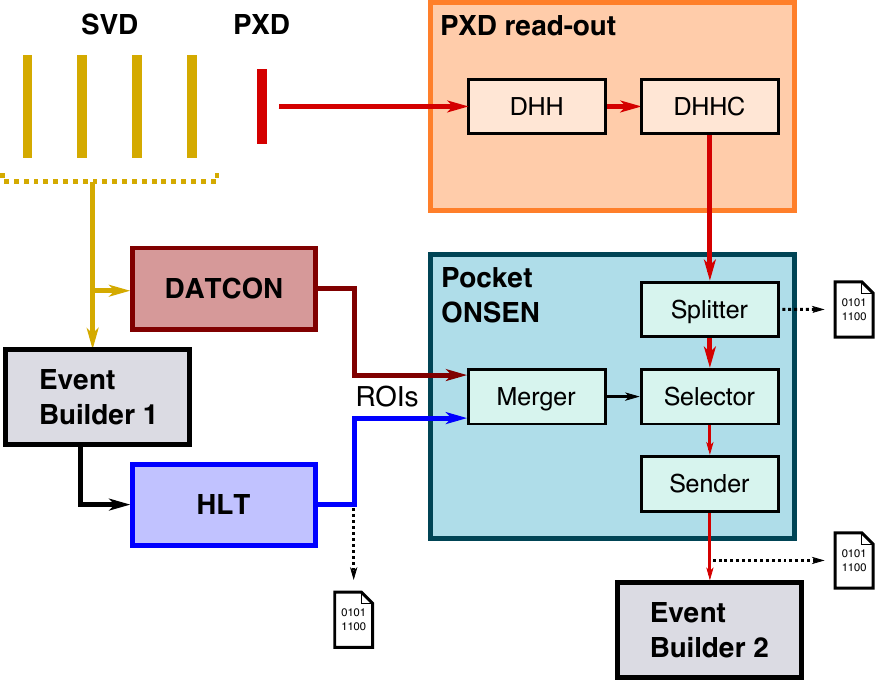}
    \caption{\label{fig:setup_tb}%
        Simplified PXD read-out and ROI generation scheme for the beam test.
        Only the subsystems relevant for PXD data reduction are shown.%
    }
\end{figure}%
Figure~\ref{fig:setup_tb} shows the data flow scheme of PXD and SVD data during
the beam test. All subsystems relevant for the data acquisition were present
(compare figure~\ref{fig:setup_belle2}). For the read-out of the single PXD
half-ladder, one DHH and one DHHC were required. The Pocket ONSEN system
included a merger node and a selector node. Both modules had all core functions
required in the final Belle~II system, including: I/O, memory access, pointer
handling, pixel data reduction, and basic slow control. Two additional modules
were used in this test:
\begin{enumerate}
    \item
    An optional \emph{splitter} node was inserted in the data path between the
    DHHC and the selector node. It forwarded the pixel data to the selector
    node and, in addition, forked it to a gigabit Ethernet output. A PC could
    be used to receive this data stream and save it to a file.
    \item
    The extensive use of debugging features in the selector nodes (mainly
    integrated logic analyzer cores) reduced the available FPGA resources. The
    considerable size of the SiTCP core therefore made it necessary to offload
    the gigabit Ether\-net output of the reduced pixel data from the selector
    to an additional \emph{sender} node. From selector to sender node, data was
    transferred via optical fiber.
\end{enumerate}
The file forked from the splitter node contained the complete PXD output data,
packed into a convenient data format by the PXD read-out system. Since this data
was yet untouched by the actual data reduction mechanism, the file could be used
for offline verification of the ONSEN system. In a similar fashion, the data
output after the sender module (i.e., the PXD data \emph{after} data reduction)
and the output from the HLT (i.e., software trigger and ROI packets) were saved
by inserting a PC in the Ethernet data stream between HLT, Pocket ONSEN, and
Event Builder 2. That way, all input and output data streams of the ONSEN system
(for runs in which the DATCON was not used) were recorded. By analyzing these
files, it was possible to check whether the PXD hits were correctly forwarded or
discarded by the ONSEN system.

A first connection test of the data acquisition chain included only the HLT,
Pocket ONSEN, and Event Builder 2. Random triggers were generated at varying
rates, for which the HLT created software trigger packets. It was verified that
the merger and selector modules in the ONSEN system processed these packets
correctly: Without PXD data for the triggers, the data from the HLT were packed
in the ONSEN output format and passed through to the Event Builder 2, which was
able to parse them and extract the correct trigger number. Besides validating
the data transport mechanism and format consistency between the various DAQ
subsystems, it could be shown that trigger rates of
\SI[per-mode=symbol]{8}{\kilo\hertz}---more than the available test beam
rates---could be handled by all systems. Note that this limit was not imposed
by the ONSEN system; previous laboratory tests, using ``dummy'' data and
trigger inputs, confirmed that the ONSEN system can cope with event rates above
the final Belle~II average trigger rate of
\SI[per-mode=symbol]{30}{\kilo\hertz}.

\begin{figure}
    \centering%
    \subfloat[][]{%
        \includegraphics{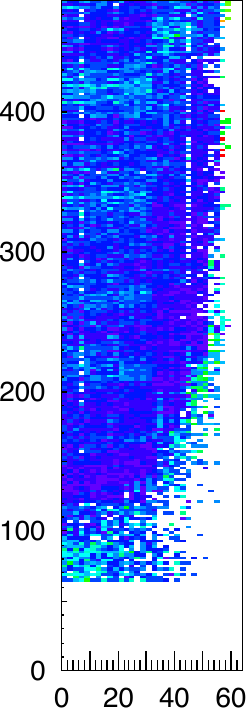}%
        \label{fig:sel_test_uncut}}%
    \hfill%
    \subfloat[][]{%
        \includegraphics{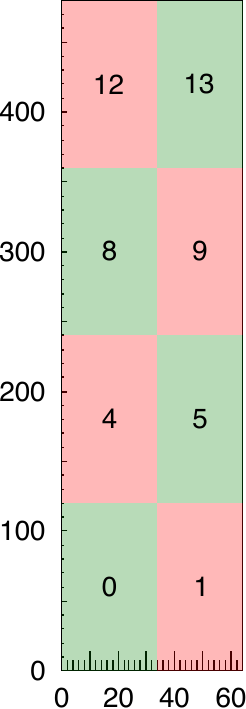}%
        \label{fig:sel_test_pattern}}%
    \hfill%
    \subfloat[][]{%
        \includegraphics{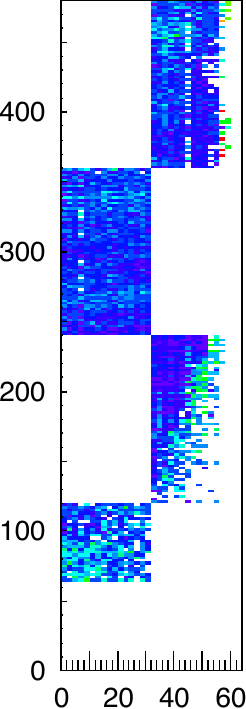}%
        \label{fig:sel_test_cut}}%
    \caption{\label{fig:sel_test}%
        ROI selection test as sanity check for the ONSEN system.
        \protect\subref{fig:sel_test_uncut}
        Hit map of data collected during a ``noise run'' (no beam, no pedestal
        configuration).
        \protect\subref{fig:sel_test_pattern}
        ROI pattern during this run: Rectangular ROIs covered different areas
        of the PXD, changing from event to event. The shaded areas were
        selected as ROIs if event number modulo 16 equaled the number shown in
        the plot.
        \protect\subref{fig:sel_test_cut}
        Same hit map after a cut on 4 values of event number modulo 16.%
    }
\end{figure}
After the installation of the detectors and read-out systems in the test beam
areas, first runs were taken with fixed ROI patterns. In particular, ROIs
covering the entire PXD matrix were sent to the ONSEN system, so that no hits
were discarded. After the complete DAQ chain had been established, noise data
from the PXD (without beam and pedestal configuration) was recorded with
varying ROI patterns. Figure~\ref{fig:sel_test} shows the results from such a
run, which was used for a rough sanity check of the ONSEN system: It was shown
that the ONSEN output only included PXD hits inside areas selected by ROIs.

\begin{figure}
     \centering%
     \includegraphics{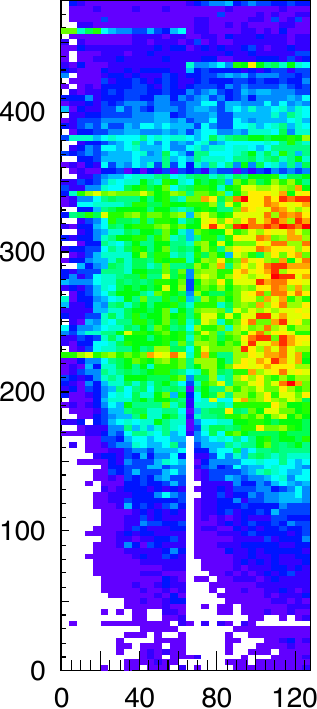}%
     \caption{\label{fig:beam_spot}%
        PXD hit map (pixel row vs.\ column) with more than 800\,000 hits from a
        beam test run with real ROIs. Hot pixels are masked out. The beam spot
        can be clearly seen.%
    }
\end{figure}

\newpage
The first data taken with beam and ``full frame'' ROIs allowed the alignment of
the various detector layers. The next step was the verification of the tracking
algorithm and generation of real ROIs~\cite{hlt_test}.
Figure~\ref{fig:beam_spot} shows a hit map of data taken with real ROIs, after
data reduction performed by the ONSEN system.

Altogether, more than 20 million events were recorded with all essential
subsystems of the DAQ chain. An analysis of the recorded ONSEN input and output
data files shows that the ONSEN system worked without errors: For all recorded
events, all pixel hits selected by ROIs were sent to the Event Builder 2, and
all other hits were discarded.

% ------------------------------------------------------------------------------
\section{Conclusion and Outlook}
The ONSEN data reduction system is an essential part of the data acquisition
system for the Belle~II pixel detector. During the combined beam test of the
Belle~II vertex detector in January 2014, we showed that the ONSEN system is in
a mature state, both in terms of hardware and FPGA firmware. The production of
the remaining modules for the system is currently ongoing. Installation will
start in 2015.

% <><><><><><><><><><><><><><><><><><><><><><><><><><><><><><><><><><><><><><><>
% Back matter
% <><><><><><><><><><><><><><><><><><><><><><><><><><><><><><><><><><><><><><><>
\appendices

% ------------------------------------------------------------------------------
\section*{Acknowledgment}
The authors would like to thank G.~Korcyl of the Jagiellonian University,
Krak{\'o}w, for providing us with his FPGA implementation of the UDP protocol
and for his continuing support in this regard.

% ------------------------------------------------------------------------------
\newpage
\bibliographystyle{IEEEtran}
\bibliography{IEEEabrv,tns}

\end{document}